%% file: eprint.tex
\documentclass[10pt]{article}
\usepackage{graphicx}
\usepackage{amsmath}
\usepackage{multirow}
\usepackage{lineno}
\usepackage{macross}
\usepackage{xspace}
\usepackage{url}

\def\Title#1{\begin{center} {\Large #1 } \end{center}}
\def\Author#1{\begin{center}{ \sc #1} \end{center}}
\def\Address#1{\begin{center}{ \it #1} \end{center}}

\newcommand\pubblock{\rightline{\begin{tabular}{l} Proceedings of the Fifth Annual LHCP\\ \pubnumber\\
         \pubdate  \end{tabular}}}

\newenvironment{Abstract}{\begin{quotation} \begin{center} 
             \large ABSTRACT \end{center}\bigskip 
      \begin{center}\begin{large}}{\end{large}\end{center} \end{quotation}}

\newenvironment{Presented}{\begin{quotation} \begin{center} 
             PRESENTED AT\end{center}\bigskip 
      \begin{center}\begin{large}}{\end{large}\end{center} \end{quotation}}

\input econfmacros.tex

\usepackage{color}

\newcommand\pubnumber{ ATL-PHYS-PROC-2017-113 }

\newcommand\pubdate{\today}

\def\affiliation{
On behalf of the ATLAS Collaboration, \\
Duke University, USA\\
Tsung-Dao Lee Institute, China\\
Shanghai Jiao Tong University, China}

\begin{document}

\large
\begin{titlepage}
\pubblock

\vfill
\Title{Monte Carlo modeling of Standard Model multi-boson production processes for $\sqrt{s} = 13$ TeV ATLAS analyses}
\vfill

\Author{ Shu Li }
\Address{\affiliation}
\vfill
\begin{Abstract}
Multi-boson production measurements provide an important test of the electroweak sector of
the Standard Model. The production of multiple gauge bosons $V$ (= $W^{\pm}$, $Z$, $\gamma$) opens up a
multitude of potential decay channels categorized according to the number of charged
leptons in the final state.
We present the Monte Carlo setup used by ATLAS to model multi-boson processes in
$\sqrt{s} = 13$ TeV proton-proton collisions. The baseline Monte Carlo generators are compared with
each other in key kinematic distributions of the processes under study.
\end{Abstract}
\vfill

\begin{Presented}
The Fifth Annual Conference\\
 on Large Hadron Collider Physics \\
Shanghai Jiao Tong University, Shanghai, China\\ 
May 15-20, 2017
\end{Presented}
\vfill
\end{titlepage}
\def\thefootnote{\fnsymbol{footnote}}
\setcounter{footnote}{0}
%

\normalsize 


\section{Motivation}

In ATLAS experiment~\cite{bib:atlas}, the multiboson productions are ones of the most important processes for both Standard Model (SM) measurements (signals) and BSM searches (backgrounds).
This study tries to investigate the performance of various benchmark generators used in ATLAS within the official software framework
and provides a summary of such process modeling performances in ATLAS.

\section{Generators}

The following setups are used for the Monte Carlo (MC) simulation:
\begin{itemize}
\item List of Generators: \sherpa~\cite{bib:sherpa}, \powhegbox~\cite{bib:powheg1,bib:powheg2,bib:powheg3}, \mg~\cite{bib:mg5}, \mcatnlo~\cite{bib:mcatnlo}, \vbfnlo~\cite{bib:vbfnlo}.
\item List of Parton Shower (PS) setups: \pythia~\cite{bib:pythia}, \herwig7/\herwig++~\cite{bib:herwig}, \sherpa~\cite{bib:sherpa}.
\end{itemize}
The modeling of the multi-jet associations in the investigated processes are done
with multi-leg matrix elements. The merging schemes being studied are CKKW, MEPS@NLO and FxFx depending on the generators~\cite{bib:pubnote}.

\section{Fully Leptonic diboson process modeling}

The fully leptonic diboson processes are modeled by various generators at high precision as shown in Table~\ref{tab:fully-accuracies}.

\begin{table}[htbp]
\centering
\caption{Overview of di-boson process accuracies for the generators. NLO: next-to-leading order, LO: leading order, PS: parton shower~\cite{bib:pubnote}.}
\label{tab:fully-accuracies}
\begin{tabular*}{\textwidth} { l c c c c c c }
 \hline
  & & $VV+0j$ & $VV+1j$ & $VV+2j$ & $VV+3j$ & $VV+\geq 4j$ \\
 \hline
 \multirow{2}{*}{$ $}         & \sherpa \texttt{v2.2}  & NLO & NLO & LO & LO & PS \\
                              & \powhegboxpythia{}/\herwig{}++    & NLO &  LO & PS & PS & PS \\
                              & \mg+\pythia      & NLO &  NLO & LO & PS & PS \\
                              & \mcatnlo+\herwig      &  NLO &  LO & PS & PS & PS \\

 \hline
\hline
\end{tabular*}
\end{table}

\begin{figure}[htb]
\centering
  \includegraphics[width=0.4\columnwidth]{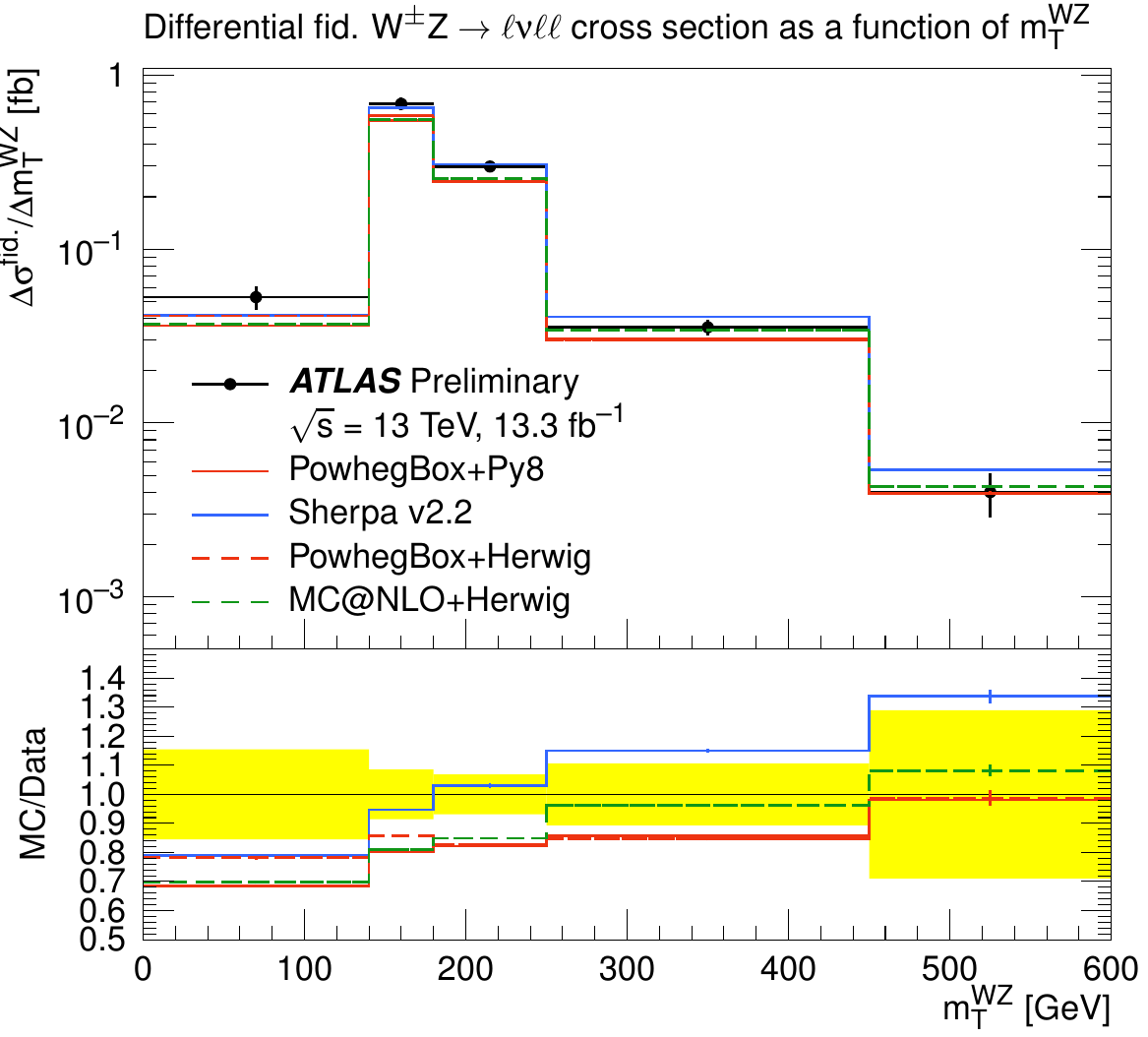}
  \includegraphics[width=0.4\columnwidth]{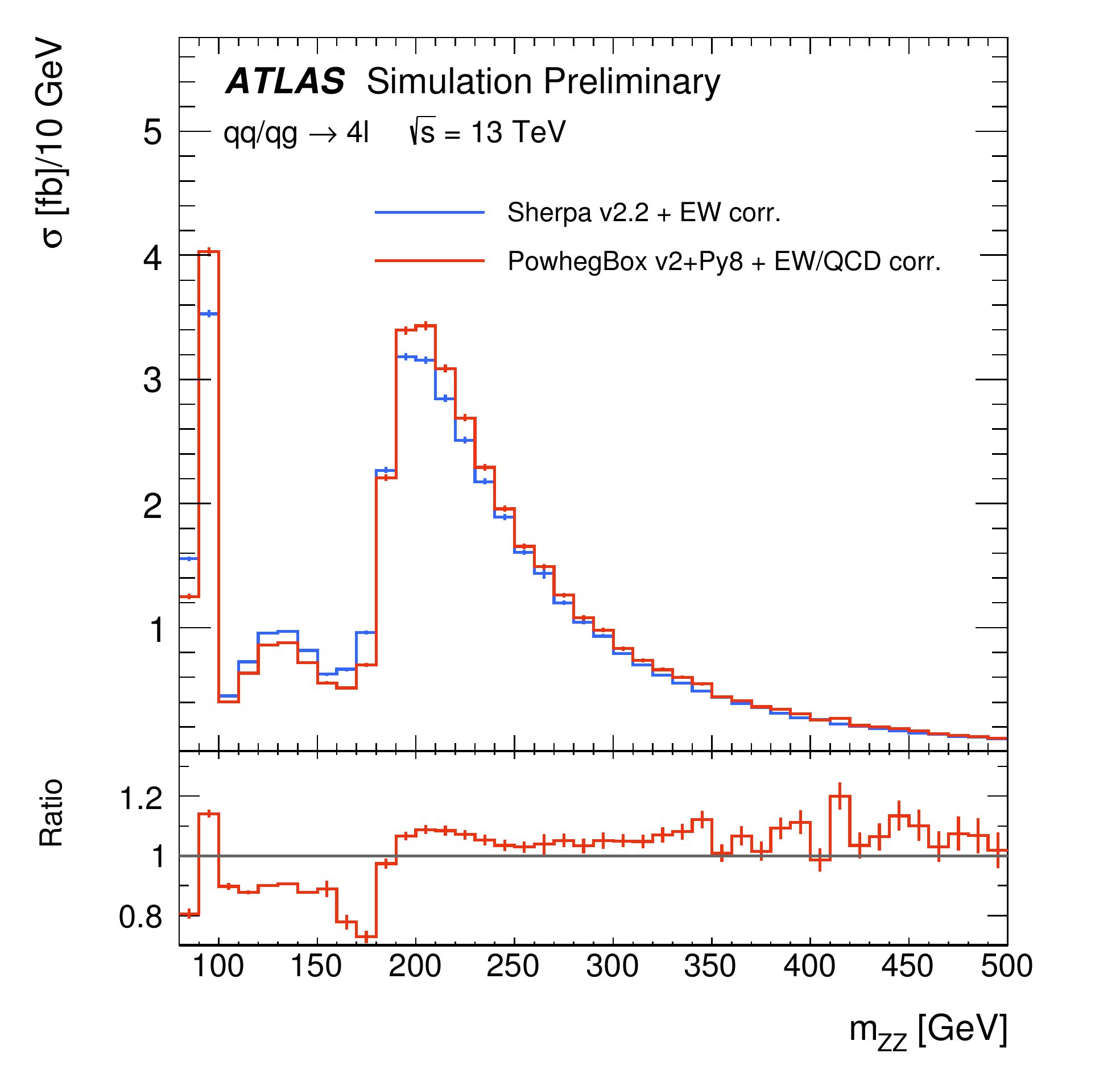}
\caption{The modeled kinematics comparison between different generators in $WZ$ and $ZZ$ processes. The plot on the left shows comparison of differential $W^{\pm}Z$ cross section as a function of
the transverse mass variable $m_{T}^{WZ}$ for the $W^{\pm}Z$ system with \powhegbox + \pythia, \powhegbox + \herwig, \sherpa and \mcatnlo + \herwig predictions.
The plot on the right shows the differential cross section predictions of \sherpa and \powhegbox both with higher-order electroweak effects and QCD effects corrections as the function of four-lepton
invariant mass, which is rather insensitive to higher-order QCD effects according to the comparison~\cite{bib:pubnote}.}
\label{fig:VV_plot}
\end{figure}

Figure~\ref{fig:VV_plot} shows the kinematics comparisons between different generators and the data measurement in $WZ$ production process and differential cross section predictions of \sherpa and \powhegbox in $ZZ$ production process.

\section{Electroweak diboson(+$jj$) process modeling}

\begin{figure}[htb]
\centering
  \includegraphics[width=0.4\columnwidth]{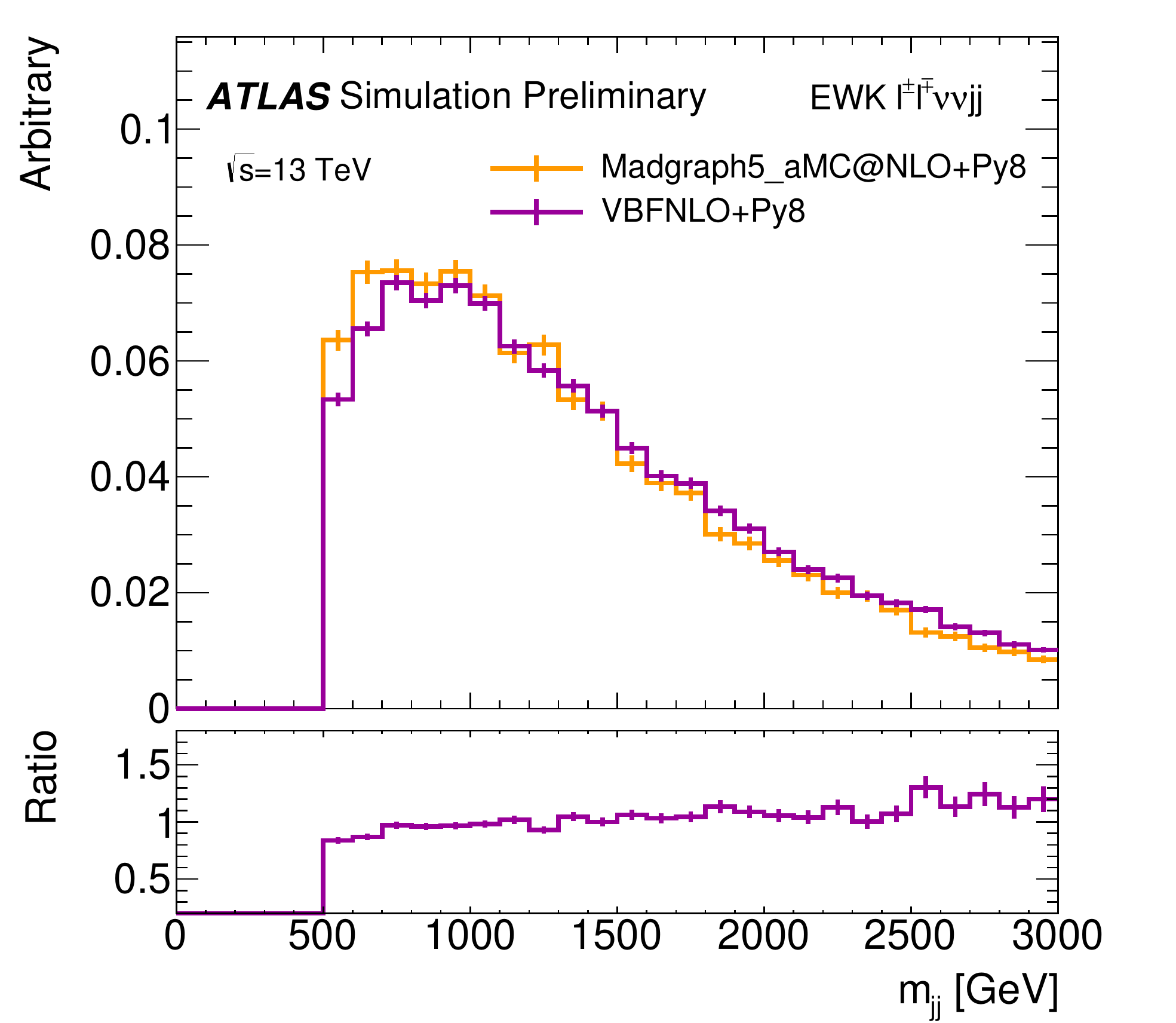}
  \includegraphics[width=0.4\columnwidth]{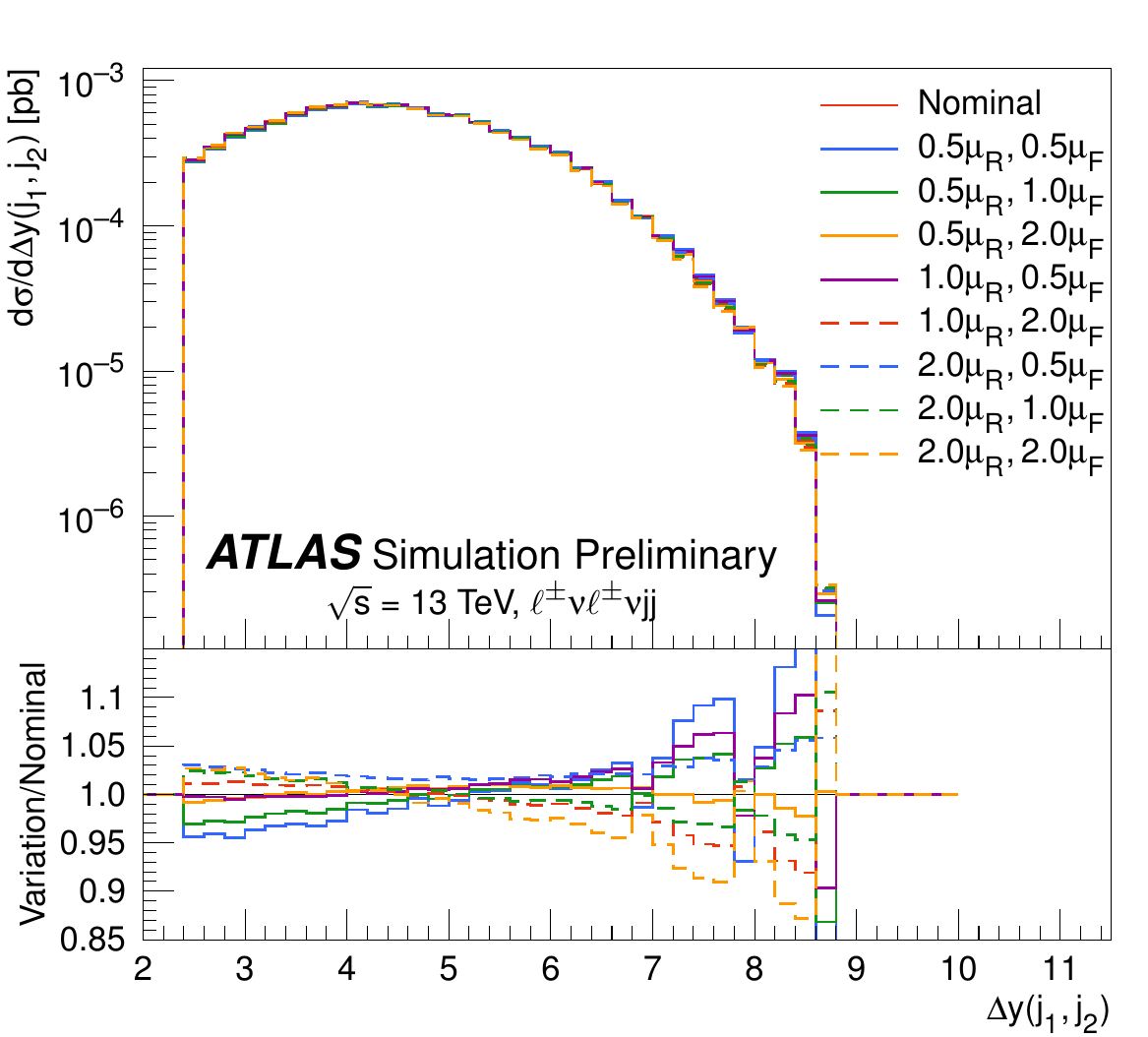}
\caption{The modeled kinematics comparison between different generators in $W^{\pm}W^{\mp}/ZZ\to\ell^{\pm}\ell^{\mp}2\nu+jj$ and with scale variations
in $W^{\pm}W^{\pm}\to\ell^{\pm}\ell^{\pm}2\nu+jj$ processes. The left plot shows in the $W^{\pm}W^{\mp}+jj$ channel the comparison of predicted kinematic distributions
between \mg and \vbfnlo, both of which are showered with \pythia, for di-jet invariant mass $m_{jj}$. The distributions is normalised to the same integral,
and the shown uncertainties are statistical only. The right plot quantifies the impact of the scale variations on the rapidity separation $\Delta y(j1,j2)$ for the two leading jets~\cite{bib:pubnote}.}
\label{fig:VVjj_plot}
\end{figure}

The modeling of most electroweak diboson production processes in association with two jets is provided with LO precision for 2-jets and higher-jet multiplicities are modeled via parton showering.
Only in the like-charged $W^{\pm}W^{\pm}\to\ell^{\pm}\ell^{\pm}2\nu+jj$ process, \powhegbox provides NLO order modeling of 2-jet bin and leading order modeling 3-jet bin.
At QCD LO without asking for vector boson decays, the electroweak (EWK) coupling order is 2 for QCD induced processes and 4 for EWK induced processes while the QCD coupling order is 2 for QCD induced processes and 0 for EWK induced processes.
Figure~\ref{fig:VVjj_plot} shows the opposite-charged $W^{\pm}W^{\mp}/ZZ\to\ell^{\pm}\ell^{\mp}2\nu+jj$ process modeling comparison between \vbfnlo and \mg, and the like-charged $W^{\pm}W^{\pm}\to\ell^{\pm}\ell^{\pm}2\nu+jj$
process modeling comparisons between different scale choices.

\section{Triboson process modeling}

\begin{figure}[htb]
\centering
  \includegraphics[width=0.4\columnwidth]{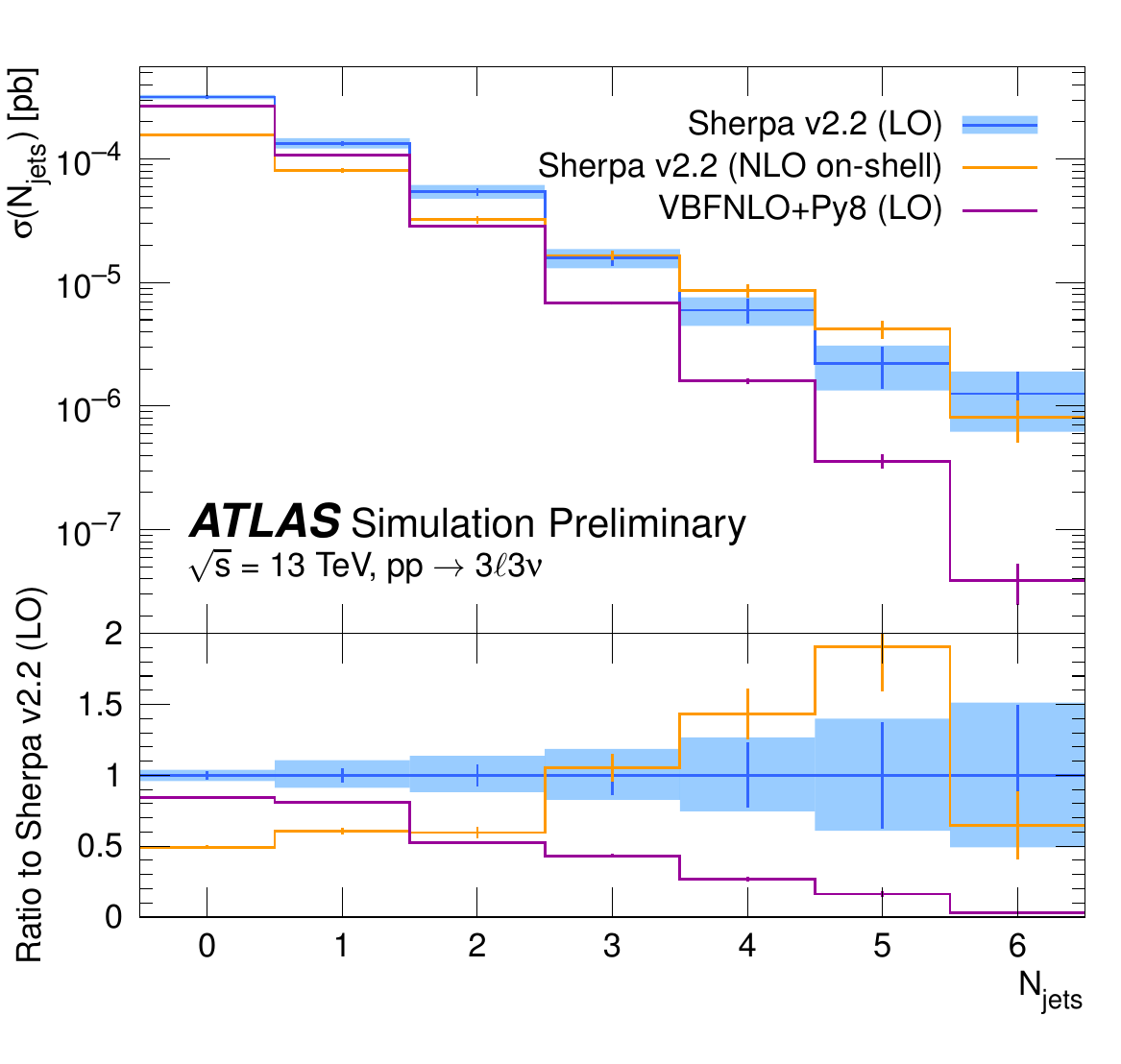}
  \includegraphics[width=0.4\columnwidth]{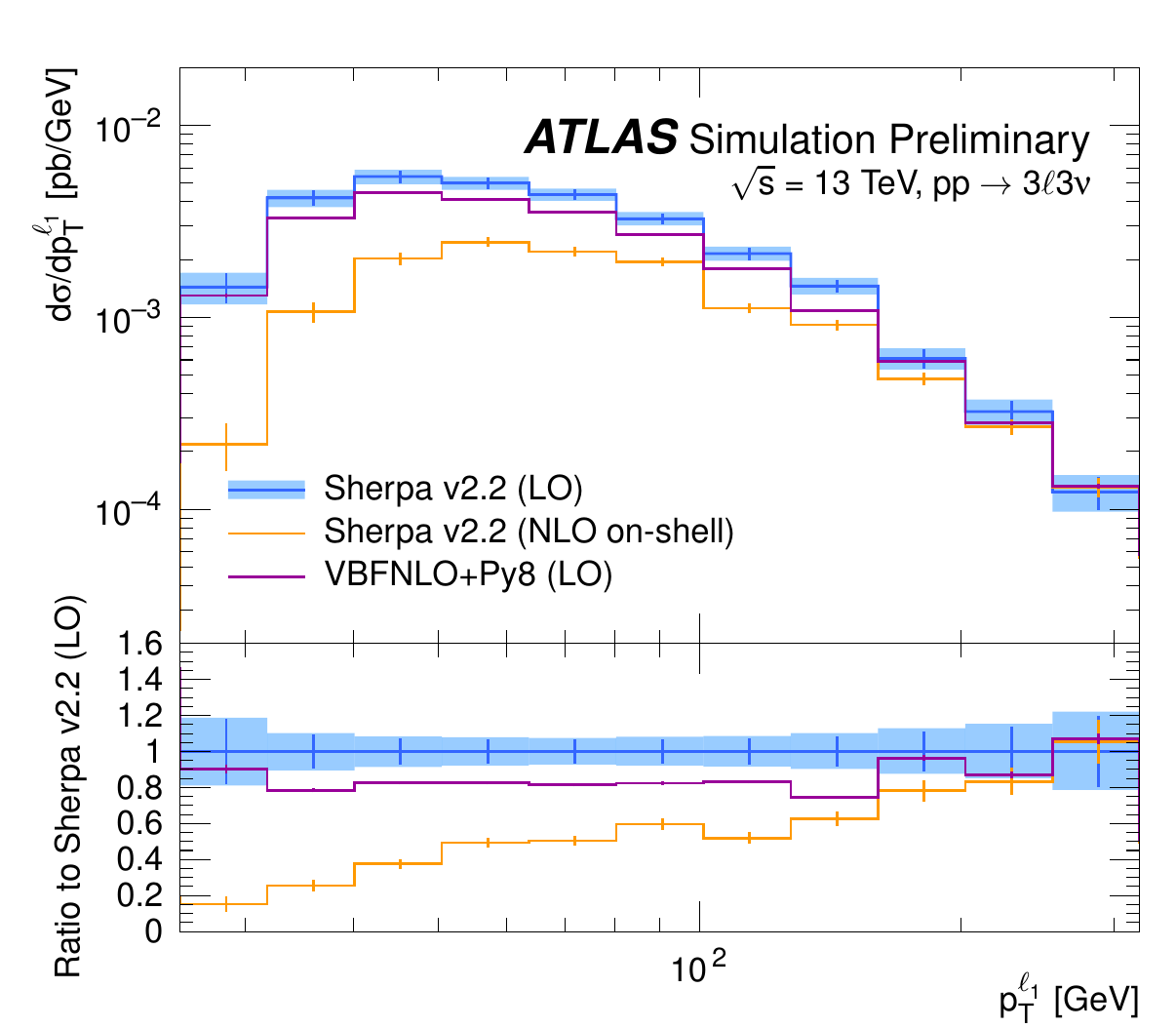}
\caption{The modeled jet multiplicity and leading lepton $p_\mathrm{T}$ comparison between different generators in $WWW\to3\ell3\nu$ process, in between \sherpa and \vbfnlo~\cite{bib:pubnote}.}
\label{fig:VVV_plot}
\end{figure}

The SM rare processes of triboson productions are modeled by \sherpa and \vbfnlo at LO. \sherpa v2.2 also provides NLO modeling of on-shell tribosons and LO modeling up to 2-jets.
Higher jet multiplicities are modeled via parton showering. Table~\ref{tab:vvv-accuracies} summarizes the modeled precisions.
Figure~\ref{fig:VVV_plot} shows the jet multiplicity and leading lepton $p_\mathrm{T}$ distribution comparisons between different generators.

\begin{table}[htbp]
\centering
\caption{Overview of triboson process accuracies for the chosen generators~\cite{bib:pubnote}.}
\label{tab:vvv-accuracies}
\begin{tabular*}{\textwidth} { l c c c c c }
 \hline
  & & $VVV+0j$ & $VVV+1j$ & $VVV+2j$ & $VVV+\geq 3j$ \\
 \hline
 \multirow{1}{*}{$VVV$ on-shell} & \sherpa \texttt{v2.2}  & NLO &  LO & LO & PS \\
 \hline
 \multirow{1}{*}{$6\ell,5\ell 1\nu, 4\ell 2\nu, 3\ell 3\nu, 2\ell 4\nu$} & \sherpa \texttt{v2.2}  & LO &  LO & PS & PS \\
 \hline
 \multirow{1}{*}{$3\ell 3\nu$} & \vbfnlopythia &  LO &  PS & PS & PS \\
\hline
\end{tabular*}
\end{table}

\clearpage

\section{Conclusion}
We present the MC process modeling of multi-boson productions used by ATLAS at 13 TeV in $pp$ collisions.
State-of-the-art generators are investigated and key kinematic distributions of the processes are compared.
Systematic uncertainties such as scale and PDF variations were also investigated and summarized in Ref.~\cite{bib:pubnote}.

\end{document}

%% file: econfmacros.tex
\def\beq{\begin{equation}}
\def\eeq#1{\label{#1}\end{equation}}
\def\eeqn{\end{equation}}

\def\beqa{\begin{eqnarray}}
\def\eeqa#1{\label{#1}\end{eqnarray}}
\def\eeqan{\end{eqnarray}}

\let\bar=\overbar

\def\W{{\cal W}}

\def\Dslash{\not{\hbox{\kern-4pt $D$}}}
\def\dslash{\not{\hbox{\kern-2pt $\del$}}}

\def\msb{{\bar{\ssstyle M \kern -1pt S}}}

